\documentclass{article}

\usepackage{amsfonts,amsmath,amsthm}
\usepackage{paralist}
\usepackage{multirow}
\usepackage{comment}
\usepackage{xspace}
\usepackage[noend,ruled]{algorithm2e}
\usepackage{xcolor}
\usepackage{tikz}
\usetikzlibrary{decorations,arrows,petri,topaths,backgrounds,shapes,positioning,fit,calc,decorations.pathreplacing,patterns,intersections}
\usepackage{subcaption}

\usepackage{authblk}

\newlength{\additionaltextwidth}
\usepackage[colorlinks,citecolor=green!60!black,linkcolor=red!60!black,pagebackref]{hyperref}
\renewcommand*{\backref}[1]{}
\renewcommand*{\backrefalt}[4]{%
\ifcase #1%
\marginpar{\tiny no cite}
\or
 $\rightarrow$~p.~#2.%
\else
  $\rightarrow$~pp.~#2.%
\fi
}

\newcommand{\N}{\ensuremath{\mathbb{N}}}

\newcommand{\electionC}{\ensuremath{C}}
\newcommand{\electionV}{\ensuremath{V}}
\newcommand{\pref}{\ensuremath{\succ}}
\newcommand{\calR}{\ensuremath{\mathcal{R}}}

\newcommand{\fpt}{\textsf{FPT}}
\newcommand{\xp}{\textsf{XP}}
\newcommand{\np}{\textsf{NP}}

\newcommand{\wone}{\textsf{W}{\ensuremath{[1]}}}
\newcommand{\wtwo}{\textsf{W}{\ensuremath{[2]}}}
\newcommand{\wt}{\textsf{W}{\ensuremath{[t]}}}
\newcommand{\Wt}[1]{\textsf{W}{\ensuremath{[#1]}}}
\newcommand{\wP}{\textsf{W}{\ensuremath{[P]}}}

\newcommand{\w}{\textsf{W}}
\newcommand{\opt}{\textsf{OPT}}
\newcommand{\Maj}{\textsf{(Maj)}}
\newcommand{\LOGSNP}{\textsf{LOGSNP}}

\newcommand{\probDodgsonScore}{\textsc{Dodgson Score}\xspace}
\newcommand{\probKemenyScore}{\textsc{Kemeny Score}\xspace}
\newcommand{\probBribery}{\textsc{Bribery}\xspace}
\newcommand{\probPricedBribery}{\textsc{$\$$Bribery}\xspace}
\newcommand{\probSwapBribery}{\textsc{Swap Bribery}\xspace}
\newcommand{\probShiftBribery}{\textsc{Shift Bribery}\xspace}
\newcommand{\probSupportBribery}{\textsc{Support Bribery}\xspace}

\newcommand{\probMajALong}{\textsc{Ma\-jor\-i\-ty\-wise Ac\-cept\-ed Bal\-lot}\xspace}

\newcommand{\probMajAShort}{\mbox{\textsc{MajAB}}\xspace}

\newcommand{\probMajA}{\probMajAShort}

\newcommand{\probVertexCover}{\textsc{Vertex Cover}\xspace}
\newcommand{\probMaxVertexCover}{\textsc{Max Vertex Cover}\xspace}
\newcommand{\probWCS}[2]{\textsc{Weighted Weft-#1 Depth-#2 Circuit Satisfiability}\xspace}
\newcommand{\WCS}[2]{\textsc{WCS(#1,#2)}\xspace}
\newcommand{\WCSMaj}[2]{\textsc{WCS(#1,#2)(Maj)}\xspace}
\newcommand{\probCCDVdAppr}{\textsc{\textsc{CCDV-$d$-Approval}}\xspace}

\newcounter{myquestionnumber}
\newcommand{\keyquestion}[1]{
  \begin{quote}
    \textbf{Key question \arabic{myquestionnumber}:}
    #1
  \end{quote}
  \stepcounter{myquestionnumber}
}

\newcommand{\mytitle}{Parameterized Algorithmics for Computational Social Choice: Nine Research Challenges}

\title{\mytitle}
\author[1]{Robert Bredereck}
\author[1]{Jiehua Chen}
\author[2]{Piotr Faliszewski}
\author[3]{Jiong Guo}
\author[1]{Rolf Niedermeier}
\author[4]{Gerhard J.\ Woeginger}

\affil[1]{Institut f\"ur Softwaretechnik und Theoretische Informatik,
  TU Berlin, Germany, \texttt{robert.bredereck@tu-berlin.de, jiehua.chen@tu-berlin.de, rolf.niedermeier@tu-berlin.de}}

\affil[2]{AGH University of Science and Technology, Krakow, Poland,
  \texttt{faliszew@agh.edu.pl}}

\affil[3]{Cluster of Excellence Multimodal Computing and Interaction, Universit\"at des Saarlandes, Saarbr\"ucken, Germany, \texttt{jguo@mmci.uni-saarland.de}}

\affil[4]{Department of Mathematics and Computer Science, TU Eindhoven, Eindhoven, The Netherlands, \texttt{gwoegi@win.tue.nl}}

\date{}

\begin{document}

\maketitle

\begin{abstract}
Computational Social Choice is an interdisciplinary research area involving
Economics, Political Science, and Social Science
on the one side, and Mathematics and Computer 
Science (including Artificial Intelligence and
Multiagent Systems) on the other side. Typical computational 
problems studied in this field
include the vulnerability of voting procedures against attacks, or preference
aggregation in multi-agent systems.
Parameterized Algorithmics is a subfield of Theoretical Computer Science 
seeking to exploit meaningful problem-specific parameters in order to 
identify tractable special cases of in general computationally hard problems.
In this paper, we propose nine of our favorite research challenges concerning the parameterized 
complexity of problems appearing in this context.
\end{abstract}

\section{Introduction}
{\bf Computational social choice}~\cite{BCE12,CELM07,Con10,ASS02,ASS10,fal-hem-hem:j:cacm-survey,Cake-book} is a relatively young interdisciplinary 
research area that brings together researchers from fields like
Artificial Intelligence,
Decision Theory, 
Discrete Mathematics,
Mathematical Economics,
Operations Research,
Political Sciences, 
Social Choice, and 
Theoretical Computer Science.
The main objective is to improve our understanding of social choice mechanisms 
and of algorithmic decision making. 
Some concrete questions belonging to this area are: 
How should a voter choose between competing political alternatives?
How should this voter rank these competing alternatives?
How should a meta-search engine aggregate many rankings into a single consensus 
ranking, reflecting an ``optimal'' compromise between various alternatives?
A long term goal is to improve decision support for decision makers that are
working in a variety of areas like electronic commerce, logistics, recommender 
systems, risk assessment, risk management, and that are fighting with massive 
data sets, large combinatorial structures, and uncertain information.

The general topic is addressed 
by a biennial \emph{International Workshop on Computational Social Choice (COMSOC)}, whose 2012 and 2014 editions have 
been held in Krak{\'o}w/Poland and in Pittsburgh/Pennsylvania/USA, respectively.
Furthermore, the topic is covered by a number of leading conferences in Artificial 
Intelligence (including \emph{AAAI}, \emph{ECAI}, \emph{IJCAI}) and by several 
specialized conferences (including \emph{AAMAS}, \emph{ADT}, \emph{EC}, \emph{SAGT}, \emph{WINE}). 
There are numerous research journals that address many aspects of computational 
social choice, including \emph{Artificial Intelligence}, 
\emph{ACM Transactions on Economics and Computation},
\emph{Autonomous Agents and Multi-Agent Systems}, 
\emph{Journal of Artificial Intelligence Research}, \emph{Mathematical Social Sciences},
\emph{Social Choice and Welfare}, and many others (including theoretical as well as 
application-oriented journals).

\medskip
{\bf Parameterized Complexity} is a branch of Theoretical Computer Science that started 
in the late 1980s and early 1990s.
The main objective is to understand computational problems with respect to multiple 
input parameters, and to classify these problems according to their inherent difficulty.
As the complexity of a problem can be measured as a function of a multitude of input 
parameters, this approach allows to classify NP-hard problems on a much finer scale 
than in classical complexity theory (where the complexity of a problem is only measured 
with respect to the number of bits in the input). 
Parameterized complexity analysis has been successfully applied 
in areas as diverse as
Algorithm Engineering,
Cognitive Sciences,
Computational Biology,
Computational Geometry,
Geographic Information Systems,
Machine Learning, and
Psychology,
to name a few.  
The first systematic work on Parameterized Complexity is 
the 1999~book \cite{DF99} by 
Downey and Fellows.

\medskip
{\bf This article} suggests future research directions and open problems in the 
intersection area between computational social choice and parameterized complexity.
Parts of this intersection area have been surveyed in a recent paper on the
parameterized complexity analysis of voting problems~\cite{BBCN12}.
The field holds numerous exciting challenges for researchers on algorithms and complexity.






\section{Preliminaries}




We summarize some basic concepts and definitions that will be used throughout this paper.

\subsection{Voting}
\paragraph{Elections.}
An election~$E:=(\electionC,\electionV)$ consists of a set $\electionC$ of $m$ alternatives~$c_1,c_2,\ldots,c_m$ 
and a list $\electionV$ of $n$~voters~$v_1,v_2,\ldots, v_n$. 
Each voter~$v$ has a linear order~$\pref_{v}$ over the set~$\electionC$ which we call a \emph{preference order}. 
For example, let~$C=\{c_1,c_2,c_3\}$ be a set of alternatives.
The preference order~$c_1\pref_{v} c_2 \pref_{v} c_3$ of voter~$v$ indicates that
$v$ likes $c_1$ most~(the $1$st position), then~$c_2$, 
and  $c_3$ least~(the $3$rd position).
For any two distinct alternatives~$c$ and $c'$, we write $c\pref_v c'$ if voter~$v$ prefers~$c$ over $c'$.
We also use the notation~$v\in \electionV$ to indicate that a voter~$v$ is in the list~$\electionV$.
By $|V|$, we mean the number of voters in $V$.

\paragraph{Voting rules.}
A voting rule~$\calR$ is a function 
that maps an election to a subset of alternatives, the set of winners.
One prominent voting rule dates back to de Condorcet~\cite{dC85}.
It selects an alternative as a (unique) winner if it beats any other alternative in head-to-head contests.
Formally speaking, given an election~$E=(\electionC, \electionV)$, an alternative~$c\in \electionC$ is 
a \emph{Condorcet winner} if any other alternative~$c'\in \electionC \setminus \{c\}$ satisfies 
\[
|\{v\in \electionV \mid c\pref_v c' \}| > |\{v \in \electionV \mid c' \pref_v c\}|\text{.}
\]
It is easy to see that there are elections for which no alternative is a
Condorcet winner. However, if a voting rule guarantees that a
Condorcet winner is elected whenever there is one, then we say that
this rule is Condorcet-consistent. (Such rules include, e.g., those
of Copeland~\cite{Goo54,BF02}, of Dodgson~\cite{Dod76}, of Kemeny~\cite{Kem59,Lev75}, and many others.)

Scoring protocols form another well-studied class of voting rules. A
scoring protocol for $m$ alternatives is defined through a vector
$(\alpha_1, \ldots, \alpha_m)$ of integers, $\alpha_1 \geq \cdots \geq
\alpha_m \geq 0$. An alternative receives $\alpha_i$ points for each voter
that ranks this alternative as $i$th best. The best known examples of
(families of) scoring rules include the Plurality rule (defined through
a vector~$(1,0,\ldots,0)$), $d$-Approval (defined through vectors of
$d$ ones followed by zeros), and the Borda rule (defined through vectors
$(m-1, m-2, \ldots, 1, 0)$).

\subsection{Parameterized Complexity}
\label{subsec:parcomp}
We assume familiarity with basic Computational Complexity Theory~\cite{AB09,GJ79} 
and only provide some definitions with respect to parameterized complexity analysis~\cite{DF13,FG06,Nie06}.

Let $\Sigma$ be an alphabet.
A \emph{parameterized problem} over~$\Sigma$ is a \emph{language}~$L\subseteq \Sigma^{*}\times \Sigma^*$.
The second part of the problem is called the \emph{parameter}.
Usually, this parameter is a non-negative integer or a tuple of non-negative integers.
For instance, an obvious parameter in voting is the number of alternatives. 
Thus, typically $L \subseteq \Sigma^*\times \N$, 
where a combined parameter can be interpreted as the maximum of its integer components.

In the following, let $L$ be a parameterized problem.
\paragraph{Fixed-parameter tractability.}
We say that $L$ is \emph{fixed-parameter tractable} or in $\fpt$
if there is an algorithm that,
given an input~$(I,k)$, 
decides in $f(k)\cdot |I|^{O(1)}$ time whether $(I,k)\in L$.
Herein $f\colon \N \to \N$ is a computable function depending only on $k$. 

\emph{Kernelization} is an alternative way of showing fixed-parameter tractability~\cite{Bod09,GN07,LMS12}.
We say that $L$ has a \emph{problem kernel} 
if there is a polynomial-time algorithm (that is, a kernelization) 
that given an instance~$(I,k)$, 
computes an equivalent instance~$(I',k')$ whose size is upper-bounded by a function in $k$, that is,
\begin{enumerate}[(i)]
\item $(I,k)$ is a yes-instance if and only if $(I',k')$ is a yes-instance, and
\item $|(I',k')|\le f(k)$ with $f\colon \N \to \N$.
\end{enumerate}
Typically, kernelizations are based on 
\emph{data reduction rules} executable in polynomial time
that help shrinking the instance size.
It is known that $L$ has a problem kernel if and only if $L$ is in 
$\fpt$~\cite{CCDF97}. 
Further, if the function~$f$ in the above kernelization definition is polynomial, 
then we also say that $L$ has a \emph{polynomial-size problem kernel}.

\paragraph{Parameterized intractability.}
As a central tool for classifying problems, Parameterized Complexity Theory provides the 
\emph{$\w$-hierarchy} consisting of the following classes and interrelations~\cite{CM08,DT11}:
\[
\fpt \subseteq \wone \subseteq \wtwo  \subseteq \cdots \subseteq \wt \subseteq \ldots \subseteq \xp\text{.}
\]
To show $\wt$-hardness for any positive integer~$t$, we use the concept of \emph{parameterized reduction}. 
Let $L, L'$ be two parameterized problems.
A \emph{parameterized reduction} from $L$ to $L'$ consists of two functions~$f \colon \Sigma^* \to \Sigma^*$
and $g \colon \N\to\N$ such that for any given instance~$(I_{L},k)$ of $L$, it holds that
\begin{enumerate}[(i)]
\item $(I_L,k)$ is a yes-instance for $L$ if and only if $(f(I_L),g(k))$ is a yes-instance for $L'$, and
\item $f$ is computable in $\fpt$ time for parameter~$k$, that is, in $h(k) \cdot |I_L|^{O(1)}$ time.
\end{enumerate}
Problem~$L'$ is $\wt$-hard if for any problem~$L$ in $\wt$, there is a parameterized reduction from $L$ to $L'$.
Typically, to show that some problem $L'$ is $\wt$-hard,
we start from some known $\wt$-hard problem $L$ and reduce $L$ to
$L'$.


\section{Nine Challenges}
In this main section of our paper, we describe nine challenges (indeed, mostly 
rather problem areas) that found our specific interest and that shall help to 
stimulate fruitful research on the parameterized complexity of computational 
social choice problems.

\subsection{ILP-Based Fixed-Parameter Tractability}
Our first challenge relates to a method of establishing fixed-parameter 
tractability results.
Let us explain this with the help of one of the most obvious parameters
in the context of voting---the number of alternatives. For many contexts
(for example, political or committee voting) it is natural to assume
that the number of alternatives is small (particularly when compared 
to the number of voters). Hence, it is important to determine the 
computational complexity of various voting problems for the case where the input 
contains only few alternatives. Indeed, there is a number of fixed-parameter tractability results in terms of the 
parameter ``number~$m$ of alternatives'' in the voting context, but
 many of them rely on a deep result from 
combinatorial optimization due to Lenstra~\cite{Len83}.
Moreover, since this result (on Integer Linear Programming) is mainly of  
theoretical interest, this may render corresponding fixed-parameter 
tractability results to be of classification nature only.
Fixed-parameter tractability results based on Integer Linear Programming
also tend to give less insight into the structural properties of the problems
than combinatorial algorithms.
The challenge we pose relates to improving this
situation by replacing integer linear programs with direct combinatorial
fixed-parameter algorithms.

Integer Linear Programming (ILP) 
is a strong classification tool for showing fixed-parameter 
tractability~\cite{Nie06}.
More specifically, Lenstra's famous 
result~\cite{Len83} (see the literature~\cite{FT87,Kan87} for some 
moderate later running time improvements) 
implies that a problem
is fixed-parameter tractable if it can be solved by an integer linear program
where the number of variables is upper-bounded by 
a function solely depending on
the considered parameter.

Perhaps the first example for such an ``ILP-based'' fixed-parameter
tractability result in the context of computational social choice
was implicitly given by Bartholdi III et al.~\cite{BTT89b} and later
improved by McCabe-Dansted~\cite{McC06}.
They developed an integer linear program to
solve the \np-hard voting problem \probDodgsonScore and gave a running time 
bound based on Lenstra's result. 
Without having explicitly stated this in their publication, this result yields 
fixed-parameter tractability
for \probDodgsonScore with respect to the parameter 
number~$m$ of alternatives.

Before coming to their integer linear program formulation for \probDodgsonScore, 
we start with a brief definition of Dodgson voting. 
The input of \probDodgsonScore is an election $E=(\electionC, \electionV)$, 
a distinguished alternative~$c\in \electionC$, and an integer~$k$. 
The question is whether one can make~$c$ the Condorcet winner by swapping a total 
number of at most~$k$ pairs of neighboring alternatives in the voters' preference 
orders. Refer to the literature~\cite{BGN10,BTT89b,HHR97} for more on the 
computational complexity status of \probDodgsonScore. 

Bartholdi III et al.~\cite{BTT89b}'s integer linear program for \probDodgsonScore reads as follows. 
It computes the Dodgson score of alternative~$c$, which is the minimum number of 
swaps of neighboring pairs of alternatives in order to make $c$ the Condorcet winner.
\begin{align*}
 & \min \sum_{i,j} j \cdot x_{i,j} \text{ subject to}\\
 & ~~~~~\forall i \in \tilde{V}: \sum_j x_{i,j} = N_i \\
 & ~~~~~\forall y \in C: \sum_{i,j} e_{i,j,y} \cdot x_{i,j} \geq d_y\\
 & ~~~~~x_{i,j} \geq 0
\end{align*}

In the above integer linear program, 
  $\tilde{V}$~denotes the set of preference order types (that is, the set of 
  \emph{different} preference orders in the given election),
  $N_i$~denotes the number of voters of type~$i$,
  $x_{i,j}$~denotes the number of voters with preference order of type~$i$ for which
  alternative~$c$ will be moved upwards by~$j$ positions,
  $e_{i,j,y}$ is~$1$ if the result of moving alternative~$c$ by~$j$ positions 
  upward
  in a preference order of type~$i$ is that $c$~gains an additional voter support against
  alternative~$y$, and $0$ otherwise.
  Furthermore, $d_y$~is the deficit of~$c$ with respect to alternative~$y$, 
  that is,
  the minimum number of voter supports that $c$~must gain against~$y$ to defeat~$y$ in a
  pairwise comparison.
  If $c$~already defeats~$y$, then~$d_y = 0$.
  See Bartholdi III et al.~\cite{BTT89b} for more details.
  Altogether, the integer linear program contains at most~$m \cdot m!$ variables~$x_{i,j}$,
where $m$~denotes the number of alternatives.
Thus, the number of variables in the described integer linear program is upper-bounded by a 
function in parameter~$m$ yielding fixed-parameter tractability due to
Lenstra's result.

Although solvability by an integer linear program with a bounded number of variables
implies fixed-parameter tractability, there is by far no guarantee for
practically efficient algorithms.
Indeed, due to a huge exponential function in the number of variables being part of
the running time bound, the above fixed-parameter tractability 
result is basically for classification.

There are numerous further 
results~\cite{ABCKNW13,BHN09,BCHKNSW14,BNW11,DS12,FHHR09b} for voting 
problems with 
the parameter number of alternatives where so far only 
using Lenstra's result leads to fixed-parameter tractability.
In summary, this gives the following research challenge. 

\keyquestion{Can the mentioned ILP-based fixed-parameter tractability results be replaced by direct combinatorial (avoiding ILPs) fixed-parameter algorithms?}

Interestingly, the \textsc{Kemeny Score} problem (refer to Section~\ref{ssec:partial-kernel} for the definition) is known to be solvable in 
$O(2^m \cdot m^2 \cdot n)$ time due to a dynamic programming algorithm, that is, 
there is a direct combinatorial fixed-parameter algorithm~\cite{BFGNR09}.
Nonetheless, in practical instances using an Integer Linear Programming
formulation with $O(m^2)$ binary variables is much more efficient, although
its theoretical running time is significantly worse~\cite{BBN14}.

\subsection{Parameterized Complexity of Bribery Problems}

Our second challenge regards the parameterized complexity of election
bribery problems under some voting rule~$\calR$. 
There are several families of bribery problems but,
generally speaking, their main idea is as follows.  
We are given some election $E = (C,V)$, some preferred
alternative $p$, and some budget $B$.  If we choose to bribe some voter $v$,
then we can modify $v$'s preference order, but we have to pay for it
(in some variants of the bribery problem the cost depends on how we
change the preference order, whereas in others the cost is fixed irrespective of
the extent to which we modify the preference order). The goal is to compute which
voters to bribe---and how to modify their preference orders---so that $p$ becomes
an $\calR$-winner and the bribery cost is at most $B$.

The study of the computational complexity of election bribery was
initiated by Faliszewski et al.~\cite{FHH09} (also refer to the work of Faliszewski and Rothe~\cite{fal-rot:b:handbook-comsoc-control-and-bribery} for a survey).  In particular, they studied the
following two problems for some voting rule~$\calR$:
\begin{itemize}
\item In the $\calR$-\probBribery problem, each voter has the same price
  (unit cost) for being bribed. In effect, we ask if it is possible
  to ensure the preferred alternative's victory by bribing at most $B$
  voters.
\item In the $\calR$-\probPricedBribery problem, each voter $v$ has an
  individual price $\pi_v$ for being bribed.
\end{itemize}
Note that in both problems the cost of bribing a voter does not
depend on how the voter's preference order is changed: The briber pays
for the ability to change the preference order in any convenient way.

Later, Elkind et al.~\cite{EFS09} introduced another
variant of the bribery problem, which they called
$\calR$-\probSwapBribery, where the cost of bribing a voter depends on
the extent to which we modify this voter's preference order. Formally,
they required that each voter $v$ has a swap-bribery price function,
which for each two alternatives $a$ and $b$ gives the cost of convincing
$v$ to swap $a$ and $b$ in the preference order (provided that $a$ and
$b$ are adjacent in this preference order at the time). To bribe a
voter $v$, we provide a sequence of pairs of alternatives to swap, the
voter looks at these pairs one by one and for each pair swaps the alternatives in the
preference order (provided they are adjacent at the time) and charges
us according to the swap-bribery function.  
Elkind et al.~\cite{EFS09} also defined a simpler variant of
\probSwapBribery, called \probShiftBribery, where all the swaps have
to involve the preferred alternative.
While both \probSwapBribery and
\probShiftBribery tend to be $\np$-complete for typical voting rules,
\probShiftBribery is much easier to solve approximately~\cite{EF10}.

For most typical voting rules, these bribery problems are
$\np$-complete.
A few polynomial-time algorithms exist only for the simplest of
voting rules, such as Plurality, or $d$-Approval and Bucklin, for the case of \probShiftBribery.
However, from the point of
view of parameterized complexity theory, there is a huge difference
between the $\calR$-\probBribery problems, where each voter has unit
cost for being bribed, and the other flavors of bribery, where each
voter has individually specified price. Indeed, when we use the number
of alternatives as the parameter, for typical voting rules~$\calR$ we
have that $\calR$-\probBribery is in $\fpt$ (this is implicit in the
work of Faliszewski et al.~\cite{FHH09}),
whereas the other types of bribery are in $\xp$~\cite{FHH09,EFS09} and it
is not known if they are in $\fpt$ or hard for $\wone$,
$\wtwo$, or some further class in the $\w$-hierarchy.  Thus, we have
the following challenge.

\keyquestion{For each typical voting rule $\calR$: what is
  the exact parameterized complexity of problems $\calR$-\probPricedBribery,
  $\calR$-\probSwapBribery, and $\calR$-\probShiftBribery when
  parameterized by the number of alternatives?}

Interestingly, this challenge is very closely related to the previous
one. For many typical voting rules~$\calR$, the proof that
$\calR$-\probBribery is in $\fpt$ uses an ILP-based algorithm. Having
such an algorithm (instead of a combinatorial one) gives very limited
insight into the nature of the problem and, thus, we cannot use the
ideas regarding $\calR$-\probBribery to build algorithms for, say,
$\calR$-\probPricedBribery. On the contrary, there
is no obvious way to extend the ILP-based algorithms to the variants
of bribery where voters have individual prices. The general idea
behind these ILP algorithms is to consider groups of voters with the
same preference orders jointly, but this does not lead to correct
solutions when these voters have differing prices.

Naturally, some authors have already studied parameterized complexity
of various types of bribery problems~\cite{DS12,BCFNN14}. However,
their results regard either parameters other than the number of
alternatives, or special pricing schemes where voters can be treated as
unified groups and ILP approaches work.

\subsection{FPT Approximation Algorithms}

In the previous challenge we asked about the exact complexity of
various election bribery problems parameterized by the number of
alternatives. However, instead of seeking exact complexity results it
might be easier to find good $\fpt$ approximation algorithms~(refer to the work of Marx~\cite{Mar06} for a survey).  
After
all, if the bribery problems turn out to be, say, $\wone$- or
$\wtwo$-hard, the approximation algorithms would give us some means of
solving them. Even if the problems turned out to be in $\fpt$, it is likely
that the approximation algorithms would be much faster in practice.
Further, even though for some of these problems there are good
polynomial-time approximation algorithms (for example, this is the
case for \probShiftBribery under positional scoring rules~\cite{EF10}), it might be
the case that $\fpt$ approximation algorithms would yield better
approximation ratios.

To see that $\fpt$ approximation algorithms can indeed outperform
polynomial-time ones in terms of achievable approximation ratios, let
us consider the \probMaxVertexCover problem (also known as \textsc{Partial Vertex Cover}). 
In this problem we are given an undirected graph $G$ and an integer~$k$. 
The goal is to pick $k$ vertices that jointly \emph{cover} as many edges as possible.
Note that a vertex \emph{cover}s all its incident edges.  As a generalization of
\probVertexCover, the problem is $\np$-complete~\cite{GJ79}, and it is
also known to be $\wone$-complete when parameterized by
$k$~\cite{GNW07}.  As far as we know, the best polynomial-time approximation algorithm for this problem known at the moment is due to
Ageev and Sviridenko~\cite{AS99}. This algorithm achieves approximation
ratio $\frac{3}{4}$, that is, if the optimal solution covers $\opt$
edges, the algorithm of Ageev and Sviridenko guarantees to cover at
least $\frac{3}{4}\opt$ edges. Marx~\cite{Mar06}, however,
has shown an algorithm that for each positive number $\varepsilon$
finds a solution for \probMaxVertexCover that covers at least
$(1-\varepsilon)\opt$ edges and runs in $\fpt$ time with respect to
parameters $k$ and $\varepsilon$.
In other words, Marx has given an $\fpt$ approximation scheme for the problem.  
Recently, this result
of Marx was generalized by Skowron and Faliszewski~\cite{SF13}.
Interestingly, the motivation for their work came from a
computational social choice context.

\newpage
This discussion motivates the following research challenge:

\keyquestion{For which computationally hard election-related problems
are there $\fpt$ approximation schemes?}

As we have indicated, bribery problems are perhaps the most natural
ones to consider here. Indeed, Bredereck et al.~\cite{BCFNN14} gave a fairly
general $\fpt$ approximation scheme for \probShiftBribery
parameterized by the number of voters, and an $\fpt$ approximation
scheme for \probShiftBribery parameterized by the number of
alternatives, but for a somewhat restrictive model of bribery costs.
The existence of an $\fpt$ approximation scheme that does not rely on
assumptions about the pricing model remains open.
Schlotter et al.~\cite{SFE11} provided an $\fpt$ approximation scheme for yet
another bribery problem, called \probSupportBribery.  Interestingly,
we are not aware of any $\fpt$ approximation algorithms for the
\probBribery and \probPricedBribery families of problems.

There are also many other types of election problems one could try to
seek $\fpt$ approximation algorithms for. For example, in control
problems we are given an election $E = (C,V)$, a preferred 
alternative~$p$, and some means of affecting the structure of the election. For
example, we might have some set of additional alternatives or voters
that can be convinced to participate in the election, or we might have
some means to remove some of the alternatives or voters. The task is to
ensure that $p$ is a winner while modifying the election as little as
possible. Election control problems were introduced by 
Bartholdi III et al.~\cite{BTT92} (later Hemaspaandra et al.~\cite{HHR07} extended the definition to include so-called
destructive control) and were studied from the parameterized
complexity perspective by several authors~\cite{BU09,FLLZ09,LZ10,ChFaNiTa2014}.
However, at the intuitive level, it seems that for control problems it
might be easier to obtain parameterized inapproximability results. The
reason for that is that often their hardness proofs (parameterized or
not) directly rely on problems that are difficult to approximate.

\subsection{Kernelization Complexity of Voting Problems}
Kernelization is one of the major tools to prove fixed-parameter tractability results.
In the last decade, parameterized complexity theory witnessed an extremely rapid development 
of kernelization results. Many problems turn out to admit polynomial-size 
problem kernels, most of them being graph-theoretical problems~\cite{GN07,Bod09,LMS12}. 

With respect to voting problems, although many have been shown to be fixed-parameter tractable, 
only very few problems are known to admit polynomial-size problem kernels. 
The main reason for this might be 
that a voting problem usually contains as input an election 
whose size essentially is upper-bounded 
by the number~$m$ of alternatives times the number~$n$ of voters.
According to the definition of polynomial-size problem kernels, 
we need to bound both numbers by polynomial functions in the considered parameter; 
it already seems difficult to bound one out of~$n$ and~$m$ 
with a function of the other. 
Next, we use the \textsc{Constructive Control by Deleting Voters} for 
$d$-Approval (\probCCDVdAppr) problem to illustrate this difficulty
and discuss how to cope with it. 

Given an election~$(\electionC,\electionV)$ under the $d$-Approval rule, each voter assigns
one point to each of the alternatives~$c$ ranked in the top~$d$ positions of his 
preference order. We say then that this voter approves~$c$. The score~$s(c)$ of an alternative~$c$ 
is the total number of points it gains from all the voters. 
The alternative with the highest score wins the election. The \probCCDVdAppr problem asks for a list of at most~$k$ voters whose 
deletion from~$V$ makes a distinguished alternative~$p$ win the election. 
\probCCDVdAppr is \np-hard for every constant~$d\ge 3$~\cite{Lin11}.  

It is not hard to show that \probCCDVdAppr is fixed-parameter tractable with respect to the number~$k$ of deleted voters.
Observe that the alternatives~$c$ with~$s(c)<s(p)$ are \emph{irrelevant} for control by deleting voters 
since the voters approving~$p$ will never be deleted 
from~$\electionV$. Let~$I$ denote the set of all irrelevant alternatives, 
and~$R:=\electionC \setminus (I \cup \{p\})$. 
Since each alternative~$c$ in~$R$ satisfies~$s(c)\ge s(p)$, 
we have to delete at least one voter
from~$V$ who approves~$c$. By the definition of $d$-Approval, deleting~$k$ 
voters can decrease the scores of at most~$d\cdot k$ alternatives, and thus
$|R|\le d\cdot k$. Let~$V_p$ be the list of voters in~$V$ approving~$p$ and~$V_R$
the list of voters approving at least one alternative in~$R$ but not~$p$. 
Obviously, the voters deleted should be all from~$V_R$. 
With~$|R|\le d\cdot k$, we can partition~$V_R$ into at most~$O((d\cdot k)^d)$ 
many classes, 
each class containing the voters in~$V_R$ which approve the same subset of~$R$. 
A brute-force algorithm can then be applied to guess the number 
of voters to be deleted from each class, resulting in an FPT running time.   
 
With respect to deriving a polynomial problem kernel for \probCCDVdAppr, 
we further need to bound~$|I|$ and~$|V|$. 
Deleting voters who solely approve the alternatives in~$I$ 
do not help making $p$ win and can be safely ``removed'' from the voter list~$V$. 
Thus, we can assume that the voters in $V$ who approve an irrelevant alternative should also approve $p$ or at least one alternative from $R$. 
This means that $V= V_p \cup V_R$.
Therefore, the key is to bound~$|V_p|$ and~$|V_R|$. 
However,  although the number of the classes 
of voters in $V_R$ is bounded, the number of voters in each class can be unbounded. 
The same holds also for~$V_p$, even if none of the voters in~$V_p$ will be deleted. 
Due to the relation between the scores of~$p$ and of the alternatives in~$R$, we cannot
reduce~$V_p$ and~$V_R$ independently. The kernelization algorithm of Wang et al.~\cite{WYGFC13} 
performs a ``reconstruction'' approach, that is, 
it keeps only the ``essential'' part of the instance, 
which consists of particular voters in~$V_R$, and then constructs 
a new instance around this essential part with new irrelevant alternatives and 
some new voters. The decisive idea behind this approach is to restore the score 
difference between~$s(p)$ and~$s(c)$ for every~$c\in R$ after the reconstruction; 
but~$s(p)$ and~$s(c)$ are much less than in the original instance and, thus, the 
number of voters needed can be bounded by a function of~$k$. 

The kernelization algorithm of Wang et al.~\cite{WYGFC13} achieves a polynomial 
problem kernel for \probCCDVdAppr with~$d$ being a constant; note that, 
with unbounded~$d$, \probCCDVdAppr is \wtwo-hard with respect to~$k$ 
\cite{WYGFC13}. It is conceivable that this reconstruction approach could lead 
to problem kernels for other fixed-parameter tractable control problems~\cite{LZ13}. However, 
compared to the diverse general tools for deriving polynomial (even linear) problem
kernels for graph-based problems, the research on kernelization for 
such two-dimensional problems as voting problems seems little developed. Only 
very few problem kernel results for voting problems are known so far. Thus, 
to close this gap is a major challenge. 

\keyquestion{What is the kernelization complexity of fixed-parameter tractable voting problems with respect 
to the number~$m$ of alternatives, the number~$n$ of voters, or some parameter less than~$m$ or~$n$? 
Can we derive polynomial (or even linear) problem kernels for some 
voting problems with the above parameters? }

\subsection{Partial Problem Kernels}
\label{ssec:partial-kernel}
As already discussed in the previous section,
the input to a voting problem contains an election
whose size is upper-bounded by the number of alternatives times the number of voters.
In addition, the input can also contain prices per voter 
or per position for each voter's preference order.
Either the number of alternatives, the number of voters,
or the total price are promising as 
a parameter for performing a parameterized complexity analysis 
because there are applications where either of these parameters is relatively 
small when compared to the other. For instance, in aggregating the outcomes 
of different search engines with the help of a meta-search engine, the 
number of voters is (relatively) small 
but the number of alternatives is large.
In political voting, the situation should typically be the other way round.
In terms of developing kernelization results, however, the current knowledge 
indicates a certain asymmetry with respect to both parameters, ``the number~$m$ of alternatives'' and ``the number~$n$ of voters''.
While we have often positive results when exploiting 
parameter $m$, for parameter~$n$ fewer positive results are known~\cite{BBCN12}.
Indeed, this also motivates the concept of ``partial problem kernels''.

Formally, a decision problem~$L$ with input instance~$(I,k)$ is 
said to admit a \emph{partial kernel} 
if there is a computable function~$d\colon \Sigma^* \to \N$ 
such that 
\begin{enumerate}[(i)]
\item $L$ is decidable in $\fpt$ time for parameter~$d(I)$, and
\item there is a polynomial-time algorithm that given~$(I,k)$,
  computes an equivalent instance~$(I',k')$ such that 
$k'\le f(k)$ and $d(I')\le g(k)$ with $f,g:\N\to \N$.
\end{enumerate}

We will discuss this issue using the rank aggregation problem
\probKemenyScore as a concrete example.

Informally speaking, \probKemenyScore asks,
given an election~$E=(\electionC,\electionV)$ and a positive number~$k$, 
whether there is a median \emph{ranking}, 
that is, a linear order over~$C$ whose total number of 
inversions with the voters' preference orders from $E$ 
is at most~$k$.
The score of such a ranking is the total number of inversions needed.
A \emph{Kemeny ranking} of an election is a median ranking with minimum score.
An example election with four alternatives and three voters is illustrated in Figure~\ref{ex:kemeny_election}.
It has only one Kemeny ranking~$a_1\succ a_2 \succ a_3 \succ a_4$ with score four. 

\begin{figure}[t!]
\begin{align*}
      v_1 \colon & a_1 \succ a_2 \succ a_3 \succ a_4 \\
      v_2 \colon & a_1 \succ a_2 \succ a_4 \succ a_3 \\
      v_3 \colon & a_3 \succ a_2 \succ a_1 \succ a_4 
\end{align*}
\caption{An election with four alternatives and three voters. The unique Kemeny ranking is $a_1\succ a_2 \succ a_3 \succ a_4$ with score four.}
\label{ex:kemeny_election}
\end{figure}

Surprisingly, \probKemenyScore remains \np-hard when the input election~$E$ 
has only four voters~\cite{DKNS01a,DKNS01b,BBD09}. 
In particular, this implies that 
there is no hope for fixed-parameter tractability with respect to the
parameter ``number of voters''. On the contrary, it is straightforward
to observe that a simple brute-force search already yields fixed-parameter
tractability with respect to the number of alternatives~\cite{BFGNR09}.

As a matter of fact, there are further natural parameterizations of 
\probKemenyScore~\cite{BFGNR09}. In particular, one may ask 
whether \probKemenyScore becomes (more) tractable if 
the preference orders of the input election are very similar on average, 
that is, the sum over their
pairwise differences divided by the number of pairs is small. 
Let us call the smallest integer at least as large as this value parameter~$d_a$.
Developing a dynamic programming 
algorithm~\cite{BFGNR09}, it has been shown that 
\probKemenyScore is fixed-parameter tractable for parameter~$d_a$ with running time $16^{d_a}\cdot \mathrm{poly}(n+m)$ where $\mathrm{poly}$ is a polynomial function.
Based on this,
one immediately obtains a trivial problem kernel of size~$O(16^{d_a})$ for parameter~$d_a$.
A natural follow-up question concerns the existence of 
a problem kernel of smaller size or even of polynomial size for \probKemenyScore when parameterized
by~$d_a$. We only have partially positive results here. The question whether
we can upper-bound 
both the number of alternatives and the number of voters by 
a polynomial function in~$d_a$ by means of polynomial-time data reduction
remains open. It is, however, possible to upper-bound the 
number of alternatives by a linear function in~$d_a$~\cite{BBN14,Sim13}.
Since \probKemenyScore is decidable in $\fpt$ time for~$m$, 
it admits a partial kernel. 
This concept has also been used in other application contexts, some of them outside 
the field of computational social choice~\cite{BGKN11,BFRS13,BCFNN14}.

Partial problem kernels of polynomial size have been shown to be useful to improve fixed-parameter tractability results.
It is unclear, however, whether or not some 
of the partial kernelization results mentioned can be replaced 
by ``full kernelization'' results.
This leads to the following challenge.

\keyquestion{Can the known partial problem kernels be improved to (full) problem kernels with non-trivial size bounds?}

We mention in passing that, focusing on voting problems, it might help to use weights 
for voters to obtain (full) problem kernels with non-trivial size bounds; 
however, this would not be a ``plain'' kernelization (where the kernelized 
instance needs to be of ``same type'' as the input instance) in case of unweighted 
input voters and may hide computational complexity in the weight functions by modifying 
the problem.

\subsection{Parameterizations of Election Data}
\label{ssec:para-election}
Social choice theory is full of (combinatorial and algorithmic) results on voting problems.
Many of these results are centered around \emph{general} elections, where \emph{every}
ordering of the alternatives is a feasible preference for any voter and where \emph{every} combination
of alternative orderings yields a feasible election.
For instance, Arrow's impossibility theorem shows the impossibility of preference aggregation for
\emph{general} elections under certain axioms (unrestricted domain, non-dictatorship, 
Pareto efficiency, and independence of irrelevant alternatives). 
Another example is the result of Hemaspaandra et al.~\cite{HHR97} that establishes
hardness of determining the winner in Dodgson elections with \emph{general} elections.

For this reason, one research branch concentrates on the investigation of specially structured
elections, where only certain combinations of alternative orderings are feasible.
\begin{itemize}
\item
An election is \emph{single-peaked} if there exists a linear order of the alternatives such 
that any voter's preference along this order is either (i) always strictly increasing, (ii) always 
strictly decreasing, or (iii) first strictly increasing and then strictly decreasing.
The example in Figure~\ref{ex:single-peaked} shows a single-peaked election with five alternatives 
and three voters.
\item
An election is \emph{single-crossing} if there exists a linear order of the voters such that for 
any pair of alternatives along this order, either (i) all voters have the same opinion on the 
ordering of these two alternatives or (ii) there is a single spot where the voters switch from 
preferring one alternative to the other one.
\item 
An election is \emph{one-dimensional Euclidean} if there exists an embedding of alternatives and
voters into the real numbers, such that every voter prefers the nearer one of each pair of alternatives.  
\end{itemize}
It is known that Arrow's impossibility result disappears on single-peaked elections~\cite{Black1948},
single-crossing elections~\cite{Roberts1977}, and on one-dimensional Euclidean elections
(which form a special case of single-peaked and of single-crossing elections; see also the paper of
Elkind, Faliszewski, and Skowron~\cite{EFS14} for some recent discussion of elections
that are both single-peaked and single-crossing).
On the algorithmic side, Walsh~\cite{Walsh2007}, Brandt et al.~\cite{BBHH2010}, and
Faliszewski et al.~\cite{FHHR11} showed that many electoral bribery, control, and manipulation problems
that are \np-hard in the general case become tractable under single-peaked elections.


\begin{figure}[t!]
\begin{align*}
      v_1 \colon & a_1 \succ a_2 \succ a_3 \succ a_4 \succ a_5 \\
      v_2 \colon & a_3 \succ a_4 \succ a_2 \succ a_1 \succ a_5 \\
      v_3 \colon & a_3 \succ a_2 \succ a_1 \succ a_4 \succ a_5  
\end{align*}
\caption{An election with five alternatives and three voters. It is single-peaked with respect to the
orders $a_1 \succ a_2 \succ a_3 \succ a_4 \succ a_5$ and $a_4 \succ a_3 \succ a_2 \succ a_1 \succ a_5$,
and  to their reverse orders.}
\label{ex:single-peaked}
\end{figure}

The three restrictions allow natural parameterizations.
Yang and Guo~\cite{YG13} considered \emph{$k$-peaked} elections as a generalization of single-peaked elections,
every preference order can have at most $k$ peaks (that is, at most $k$ rising streaks that alternate with 
falling streaks).
Similarly, we can generalize single-crossing elections to $k$-crossing elections, where for every pair
of alternatives there are at most $k$ spots where the voters switch from preferring one alternative to the 
other one.
A natural generalization of one-dimensional Euclidean elections are $k$-dimensional Euclidean elections,
where alternatives and voters are embedded in $k$-dimensional Euclidean space.

\keyquestion{How does the complexity of the standard bribery, manipulation and control problems in election systems
depend on the parameter $k$ for (i) $k$-peaked, (ii) $k$-crossing, (iii) $k$-dimensional Euclidean elections?}

A first step in this direction would be to get a good combinatorial understanding of such elections.
The literature contains forbidden substructure characterizations for single-peaked elections~\cite{BalHae2011} and for single-crossing elections~\cite{BreCheWoe2013}.
Knoblauch~\cite{Knoblauch2010} discussed the structure of one-dimensional Euclidean elections.
Chen et al.~\cite{ChPrWo} studied minimal the relation between single-peaked and single-crossing elections and one-dimensional Euclidean elections.
Bogomolnaia and Laslier~\cite{BL07},
and Bulteau and Chen~\cite{BuCh} investigated the restrictiveness of $k$-dimensional Euclidean elections.
To our knowledge, the combinatorial structure of the above parameterized variants has not been investigated so far.

\subsection{Distance to Tractable Cases}
``Treewidth'' is a famous concept in algorithmic graph theory.
Informally speaking, it uses a natural number to express how far a given undirected graph is away from being a 
tree:
a graph with treewidth one is a tree or a forest, and the smaller the treewidth is, 
the closer the graph is to being a tree~\cite{Bod06,BK08,Die12}. 
The interest for treewidth stems from the fact that many \np-hard graph problems can be solved 
efficiently (say by greedy or dynamic programming algorithms) when restricted to trees.
Hence, it is natural to ask whether such tractability results can be extended beyond trees. 
Treewidth is one measure that turned out particularly useful for such a quest. 
More specifically, many \np-hard graph problems (including \textsc{Clique}, \textsc{Independent Set}, 
\textsc{Dominating Set}) can be solved in linear time on graphs with bounded treewidth (assuming 
that a corresponding ``tree decomposition'' is part of the input)~\cite{Bod06,BK08,DF13,Nie06}.
Put differently, these problems are fixed-parameter tractable with respect to the parameter 
``treewidth'' of the input graph.

Also in the context of voting, there exist certain structural properties of inputs that make some of the
otherwise \np-hard problems tractable.
For instance, under single-peakedness (as discussed in Section~\ref{ssec:para-election}) many 
computationally hard winner determination problems, including \probKemenyScore~\cite{BBHH2010} 
(as discussed in Section~\ref{ssec:partial-kernel}), turn out to be polynomial-time solvable.
In analogy to the step from actual trees to ``nearly'' trees (that is, graphs with bounded treewidth), the 
computational social choice community has taken the step from actual single-peakedness to 
``nearly'' single-peakedness, all in the spirit of pushing the borders of tractability.
The following notions of nearly single-peaked elections have been studied in~\cite{ELP13,CGS13,BCW13,FHH14,cor-gal-spa:c:sp-width}.
\begin{itemize}
\item Elections that can be made single-peaked by deletion of $k$~voters (also called \emph{Maverick} voters). 
\item Elections that can be made single-peaked by deletion of $k$~alternatives.
\item Elections that can be made single-peaked by $k$~swaps in the preferences of every voter.
\item Elections that can be made single-peaked by contracting groups of up to $k$~alternatives into 
single alternatives (here the alternatives of every contracted group must show up consecutively in 
the preferences of every voter).
\end{itemize}
Each of these notions introduces a distance measure (or \emph{width} measure) to the tractable 
case of single-peakedness. 
In the same spirit, Elkind et al.~\cite{EFS12}, Bredereck et al.~\cite{BCW13}, and Cornaz
et al.~\cite{CGS13} introduced 
distance measures of elections to the tractable case of single-crossingness (as discussed 
in Section~\ref{ssec:para-election}) and to the tractable case of group separability (an election is
group separable if the alternatives can be partitioned into two non-empty groups such that every
voter prefers all alternatives of one group to all alternatives of the other group).
 
Altogether, these approaches lead to natural and meaningful ``width-bounded'' measures for
preference orders which shall, in the same spirit as treewidth and other width-related 
parameters do for graphs, lead to interesting special cases allowing to obtain fixed-parameter 
tractability results for social choice problems with respect to these width parameters.
Of course, coming up with natural and meaningful ``width-bounded'' measures is an ongoing 
challenging task.
Nonetheless, considering the huge amount and great success of work on width measures 
for graphs, there appear to be many research opportunities for working on these measures 
in the context of voting and related problems.
We remark that, from the viewpoint of parameterized complexity analysis, such investigations 
would fall under the category ``distance from triviality''-parameterizations~\cite{Cai03,GHN04}.
For applications in computational social choice, this line of research is still in its infancy 
with few results~\cite{CGS13}, leading to the following generic challenge.

\keyquestion{How can natural and meaningful ``width-based'' parameters be used to gain (practically useful)
fixed-parameter tractability results for \np-hard voting problems?}

\subsection{$\w$-Hierarchy and Majority-Based Problems}
This theoretical challenge is about creating a framework that
could simplify the $\w$-hierarchy classification of problems using
majority-based properties which naturally occur in Computational Social Choice.

Problems that are presumably not fixed-parameter tractable are usually
classified in the $\w$-hierarchy which is defined via the 
\probWCS{$t$}{$d$} (\WCS{$t$}{$d$}) problem~\cite{DF13}.

The input of \WCS{$t$}{$d$} is a Boolean circuit 
(formally, these are vertex-labeled directed acyclic graphs) of \emph{depth}~$d$ and \emph{weft}~$t$, and an integer bound $k$.
The question is whether there is a satisfying assignment of weight~$k$.
In this context, Boolean circuits are allowed to contain NOT gates, small AND
and OR gates of fan-in at most two and large AND and OR gates of 
arbitrary finite fan-in.
The weft of a Boolean circuit is the maximum number of large gates on a path from an input to the output.
The depth is the maximum number of all gates on a path from an input to the
output.
The weight of an assignment is the number of variables set to TRUE.

For any $t\ge 1$, $\wt$ is the class of parameterized problems for which a
parameterized reduction to \WCS{$t$}{$d$} exists for some  constant~$d\ge 1$ (regarding the parameter~$k$). 

Classifying a parameterized problem in the \w-hierarchy consists of two parts.
The first part is to show $\wt$-membership for some integer~$t$.
This is often done by providing a parameterized reduction to \WCS{$t$}{$d$}
for some $d$ or to a parameterized problem which is already known to be in $\wt$.

The second part is to find a lower bound on the complexity, that is, to find some integer~$t'$ such that the problem is $\Wt{t'}$-hard.
This is usually done by describing a parameterized reduction from some known $\Wt{t'}$-hard problem.
In the ideal case, where $t=t'$, one precisely classifies the problem within the \w-hierarchy.
It is also possible that one cannot find a~$t$ such that the problem is in $\wt$,
because the problem is $\Wt{t'}$-hard for all $t'$ or $\wP$-hard or \LOGSNP-hard~\cite{PY96} or \np-hard for constant parameter values.

Sometimes showing \wt-membership appears to be more challenging than showing hardness.
This is due to the nature of the investigated problem.
For instance, a problem which is based on some majority properties cannot be easily formulated via Boolean circuits
containing only NOT, AND, and OR gates, because the majority operator cannot
be easily simulated via a single large AND or OR gate.

A concrete example is the \probMajALong (\probMajA) problem.
Its input is a set $\mathcal{P}$ of $m$ proposals, a society of $n$~voters with favorite
ballots $B_1,\ldots,B_n\subseteq\mathcal{P}$, and an agenda $Q_+\subseteq \mathcal{P}$.
The question is whether there is a ballot~$Q$, $Q_{+}\subseteq Q \subseteq \mathcal{P}$, such that
a \emph{strict majority} of the voters \emph{accept}s~$Q$.
A voter~$i$ accepts ballot~$Q$ if $|B_i\cap Q|>|Q|/2$.
\probMajA is $\wtwo$-hard for the parameter size~$|Q|$ of the accepted ballot even if $Q_{+}=\emptyset$~\cite{ABCKNW13}.
However, showing \wtwo-membership seems challenging.
Note that the \emph{unanimous} version of this problem, where 
one asks for a ballot~$Q\subset \mathcal{P}$ that is accepted by all voters, is \wtwo-complete.

Modifying the set of allowed gates in \WCS{$t$}{$d$} dramatically simplifies the task of
finding a membership proof but it may lead to a slightly different \w-hierarchy:
For instance,
we can show that \probMajA can be reduced to a \WCS{$2$}{$d$} problem variant,
where instead of NOT, AND, or OR gates, majority gates (which output TRUE if the majority of inputs is TRUE) are used.
This version of the problem which we call \WCSMaj{$t$}{$d$}, and the corresponding \w\Maj-hierarchy
have been studied by Fellows et al.~\cite{FFHMR10}.
While $\wone=\wone\Maj$, for $t\ge2$, it is only known that $\wt\subseteq\wt\Maj$.
It would be interesting to know whether $\wt\Maj\subseteq\wt$ also holds.
If one cannot show this, then
it would be nice to have \wt\Maj-complete problems to show intractability results.

\keyquestion{
  How does the \w\Maj-hierarchy relate to the classical \w-hierarchy?
What are accessible complete problems for the \w\Maj-hierarchy?
}

Besides majority variants of \textsc{Set Cover} and \textsc{Hitting Set} 
(see \cite[Sec.~7]{FFHMR10})
promising natural candidates like \probMajA for \wtwo(Maj)-complete problems may occur in the context
of computational social choice, where majority-based properties 
such as Condorcet winner frequently occur.

\subsection{Cake Cutting}
In the cake division problem, $n\ge2$ players want to cut a cake into $n$ pieces so 
that every player gets a ``fair'' share of the cake according to his own private measure.
The cake represents a heterogeneous divisible good, and usually is modeled as the 
unit-interval ${\cal C}=[0,1]$.
Every player $p$ ($1\le p\le n$) has his own measure $\mu_p$ on the subsets of ${\cal C}$.
These measures are assumed to be well-behaved, and to satisfy the following conditions:
\begin{itemize}
\item Positivity:~ $\mu_p(X)\ge0$ for all $\emptyset\ne X\subseteq{\cal C}$.
\item Additivity:~ $\mu_p(X)+\mu_p(X')=\mu_p(X\cup X')$ for disjoint subsets 
$X,X'\subseteq {\cal C}$.  
\item Continuity:~ For every $X\subseteq {\cal C}$ and for every $\lambda$ with 
$0\le\lambda\le1$, there exists a subset $X'\subseteq X$ such that 
$\mu_p(X')=\lambda\cdot \mu_p(X)$.
\item Normalization:~ $\mu_p({\cal C})=1$.
\end{itemize}
A cake division is \emph{proportional} if every player receives a piece that he values 
at least $1/n$; 
the division is \emph{envy-free} if every player $p$ gets a piece that he values at least 
as much as the piece of any other player; 
it is \emph{equitable} if every player $p$ gets a piece that he values exactly $1/n$.
A \emph{protocol} is a procedure that controls the division process of the cake ${\cal C}$.

The book \cite{Cake-book} by Robertson and Webb provides an excellent summary of the area, 
and in particular covers all kinds of algorithmic aspects.
There already exists a rich literature on cake cutting, and various authors have come up 
with dozens of proportional or envy-free protocols.
In most of the known protocols, the measures $\mu_p$ enter the game as a black box and 
can be queried and evaluated for given cake pieces; the standard goal then is to design 
protocols that minimize the total number of queries.
For getting algorithmic problems in our classical algorithmic sense, however, we have to 
fully specify the problem data; in particular, the measures $\mu_p$ must be given explicitly 
(and not just as a black box!) and must be represented succinctly.
A natural approach is to represent the measures through piecewise linear value density 
functions $f_p:[0,1]\to\mathbb{R}$.
A feasible piece of cake then consists of finitely many intervals, and the corresponding 
measure is the integral of $f_p$ taken over all these intervals; note that the normalization 
condition implies that the integral of $f_p$ over the full interval $[0,1]$ equals~$1$.

Brams et al.~\cite{BraJonKla2013} provided a three-player example with piecewise
linear value density functions that has (i) an envy-free division and (ii) an equitable
division, but that does not allow a division that is simultaneously envy-free and equitable.  
Kurokawa et al.~\cite{KurLaiPro2013} also considered value density functions that are piecewise linear.
They design an envy-free cake cutting protocol which requires a number of operations that is polynomial in some natural parameters of the given instance.
Aumann et al.~\cite{AumDomHas2013} considered value density functions that are 
piecewise constant, and additionally require that each player ends up with a \emph{single} 
subinterval of the cake $[0,1]$; in this model it is NP-hard to find a division that 
maximizes the utilitarian welfare, that is, the sum of the values that the $n$ players assign 
to their respective pieces.

There are some straightforward parameterizations that naturally generalize the models that 
we just discussed: 
The value density function of every player consists of at most $\alpha$ pieces, and each
such function piece is a polynomial function of degree at most $\beta$.
A feasible division of the unit-interval cake (for $n$ players) consists of at most 
$\gamma n$ subintervals altogether, and every player receives at most $\delta$ subintervals.
Instead of using the unit-interval, the cake could also be modeled as the $2$-dimensional 
unit-square, or as some simple $d$-dimensional object (with the dimension $d$ as parameter); 
the subintervals then would translate into simple polyhedral pieces with a small number of 
facets (and with a number of additional parameters for measuring simplicity).

\keyquestion{How hard is it to find proportional or envy-free or equitable divisions under the above parameterizations?}

Clearly, there is an abundance of other ways for parameterizing the cake-cutting world.
As yet another challenging objective function, we mention the egalitarian welfare, the 
minimum of the values that the $n$ players assign to their respective pieces.

%
%
\section{Conclusion}

The purpose of this work is to stimulate more research in a promising application 
field of computational complexity analysis and algorithmics.
Numerous further research challenges can be found in the very active 
and steadily growing area of computational social choice.
Clearly, our selection of topics was subjective and could be easily extended.
For instance, in voting contexts it often makes sense (motivated by 
real-world data) to deal with partial instead of linear orders, generally 
making the considered problems harder. Here, one might parameterize on 
the ``degree of non-linearity''. Additional interesting (broad) research topics 
would be to further explore on relations to fields such as scheduling or 
graph and matching theory. 
By their very nature, many problems in computational social choice 
carry many natural parameterizations, motivating a thorough 
multivariate computational complexity analysis~\cite{FJR13,Nie10}.
Finally, so far there is little 
work in terms of algorithm engineering for empirically validating and 
improving the performance of fixed-parameter algorithms for 
social choice problems.

\paragraph{Acknowledgments.} 
We thank Britta Dorn, Dominikus Kr\"uger, J\"org Rothe, Lena Schend, Arkardii Slinko, Nimrod Talmon, and an anonymous referee for their constructive feedback on previous versions of the manuscript.
Robert Bredereck and Piotr Faliszewski were supported by the Deutsche Forschungsgemeinschaft, project PAWS (NI 369/10).
Jiehua Chen was supported by the Studienstiftung des Deutschen Volkes, 
Jiong Guo was supported by DFG ``Cluster of Excellence Multimodal Computing and Interaction'', 
and Gerhard Woeginger was supported by DIAMANT (a mathematics cluster of the Netherlands Organization for
Scientific Research NWO) and by the Alexander von Humboldt Foundation, Bonn, Germany.

\bibliographystyle{abbrv}
\bibliography{bib}

\begin{thebibliography}{10}

\bibitem{AS99}
A.~A. Ageev and M.~Sviridenko.
\newblock Approximation algorithms for maximum coverage and max cut with given
  sizes of parts.
\newblock In {\em Integer Programming and Combinatorial Optimization
  (IPCO'99)}, volume 1610 of {\em LNCS}, pages 17--30. Springer, 1999.

\bibitem{ABCKNW13}
N.~Alon, R.~Bredereck, J.~Chen, S.~Kratsch, R.~Niedermeier, and G.~J.
  Woeginger.
\newblock How to put through your agenda in collective binary decisions.
\newblock In {\em Proceedings of the 3rd International Conference on
  Algorithmic Decision Theory (ADT'13)}, volume 8176 of {\em LNCS}, pages
  30--44, 2013.

\bibitem{AB09}
S.~Arora and B.~Barak.
\newblock {\em Computational Complexity: A Modern Approach}.
\newblock Cambridge University Press, 2009.

\bibitem{ASS02}
K.~J. Arrow, A.~K. Sen, and K.~Suzumura, editors.
\newblock {\em Handbook of Social Choice and Welfare, Volume 1}.
\newblock North-Holland, 2002.

\bibitem{ASS10}
K.~J. Arrow, A.~K. Sen, and K.~Suzumura, editors.
\newblock {\em Handbook of Social Choice and Welfare, Volume 2}.
\newblock North-Holland, 2010.

\bibitem{AumDomHas2013}
Y.~Aumann, Y.~Dombb, and A.~Hassidim.
\newblock Computing socially-efficient cake divisions.
\newblock In {\em Proceedings of the 12th International Conference on
  Autonomous Agents and Multiagent Systems (AAMAS'13)}, pages 343--350.
  IFAAMAS, 2013.

\bibitem{BalHae2011}
M.~A. Ballester and G.~Haeringer.
\newblock A characterization of the single-peaked domain.
\newblock {\em Social Choice and Welfare}, 36(2):305--322, 2011.

\bibitem{BTT92}
J.~{Bartholdi III}, C.~Tovey, and M.~Trick.
\newblock How hard is it to control an election?
\newblock {\em Mathematical and Computer Modelling}, 16(8/9):27--40, 1992.

\bibitem{BTT89b}
J.~J. {Bartholdi~III}, C.~A. Tovey, and M.~A. Trick.
\newblock Voting schemes for which it can be difficult to tell who won the
  election.
\newblock {\em Social Choice and Welfare}, 6(2):157--165, 1989.

\bibitem{BFRS13}
M.~Basavaraju, M.~C. Francis, M.~S. Ramanujan, and S.~Saurabh.
\newblock Partially polynomial kernels for {S}et {C}over and {T}est {C}over.
\newblock In {\em Foundations of Software Technology and Theoretical Computer
  Science (FSTTCS'13)}, volume~24 of {\em LIPIcs}, pages 67--78. Schloss
  Dagstuhl--Leibniz-Zentrum f\"ur Informatik, 2013.

\bibitem{BBCN12}
N.~Betzler, R.~Bredereck, J.~Chen, and R.~Niedermeier.
\newblock Studies in computational aspects of voting---a parameterized
  complexity perspective.
\newblock In {\em The Multivariate Algorithmic Revolution and Beyond}, volume
  7370 of {\em LNCS}, pages 318--363. Springer, 2012.

\bibitem{BBN14}
N.~Betzler, R.~Bredereck, and R.~Niedermeier.
\newblock Theoretical and empirical evaluation of data reduction for exact
  {K}emeny rank aggregation.
\newblock {\em Autonomous Agents and Multi-Agent Systems}, 28(5):721--748,
  2014.

\bibitem{BFGNR09}
N.~Betzler, M.~R. Fellows, J.~Guo, R.~Niedermeier, and F.~A. Rosamond.
\newblock Fixed-parameter algorithms for {K}emeny rankings.
\newblock {\em Theoretical Computer Science}, 410(45):4554--4570, 2009.

\bibitem{BGKN11}
N.~Betzler, J.~Guo, C.~Komusiewicz, and R.~Niedermeier.
\newblock Average parameterization and partial kernelization for computing
  medians.
\newblock {\em Journal of Computer and System Sciences}, 77(4):774--789, 2011.

\bibitem{BGN10}
N.~Betzler, J.~Guo, and R.~Niedermeier.
\newblock Parameterized computational complexity of {D}odgson and {Y}oung
  elections.
\newblock {\em Information and Computation}, 208(2):165--177, 2010.

\bibitem{BHN09}
N.~Betzler, S.~Hemmann, and R.~Niedermeier.
\newblock A multivariate complexity analysis of determining possible winners
  given incomplete votes.
\newblock In {\em Proceedings of the 21st International Joint Conference on
  Artificial Intelligence (IJCAI'09)}, pages 53--58. AAAI Press, 2009.

\bibitem{BNW11}
N.~Betzler, R.~Niedermeier, and G.~J. Woeginger.
\newblock Unweighted coalitional manipulation under the {B}orda rule is
  {NP}-hard.
\newblock In {\em Proceedings of the 22th International Joint Conference on
  Artificial Intelligence (IJCAI'11)}, pages 55--60. AAAI Press, 2011.

\bibitem{BU09}
N.~Betzler and J.~Uhlmann.
\newblock Parameterized complexity of candidate control in elections and
  related digraph problems.
\newblock {\em Theoretical Computer Science}, 410(52):43--53, 2009.

\bibitem{BBD09}
T.~Biedl, F.~J. Brandenburg, and X.~Deng.
\newblock On the complexity of crossings in permutations.
\newblock {\em Discrete Mathematics}, 309(7):1813--1823, 2009.

\bibitem{Black1948}
D.~Black.
\newblock On the rationale of group decision making.
\newblock {\em Journal of Political Economy}, 56(1):23--34, 1948.

\bibitem{Bod06}
H.~L. Bodlaender.
\newblock Treewidth: Characterizations, applications, and computations.
\newblock In {\em Proceedings of the 32nd International Workshop on
  Graph-Theoretic Concepts in Computer Science (WG'06)}, volume 4271 of {\em
  LNCS}, pages 1--14, 2006.

\bibitem{Bod09}
H.~L. Bodlaender.
\newblock Kernelization: New upper and lower bound techniques.
\newblock In {\em Proceedings of the 4th International Workshop on
  Parameterized and Exact Computation (IWPEC'09)}, volume 5917 of {\em LNCS},
  pages 17--37. Springer, 2009.

\bibitem{BK08}
H.~L. Bodlaender and A.~M. C.~A. Koster.
\newblock Combinatorial optimization on graphs of bounded treewidth.
\newblock {\em The Computer Journal}, 51(3):255--269, 2008.

\bibitem{BL07}
A.~Bogomolnaia and J.-F. Laslier.
\newblock Euclidean preferences.
\newblock {\em Journal of Mathematical Economics}, 43(2):87--98, 2007.

\bibitem{BF02}
S.~Brams and P.~C. Fishburn.
\newblock Voting procedures.
\newblock In K.~J. Arrow, A.~K. Sen, and K.~Suzumura, editors, {\em Handbook of
  Social Choice and Welfare}, volume~1, pages 173--236. Elsevier, 2002.

\bibitem{BraJonKla2013}
S.~J. Brams, M.~A. Jones, and C.~Klamler.
\newblock {\it N}-person cake-cutting: There may be no perfect division.
\newblock {\em The American Mathematical Monthly}, 120(1):35--47, 2013.

\bibitem{BBHH2010}
F.~Brandt, M.~Brill, E.~Hemaspaandra, and L.~A. Hemaspaandra.
\newblock Bypassing combinatorial protections: Polynomial-time algorithms for
  single-peaked electorates.
\newblock In {\em Proceedings of the 24th AAAI Conference on Artificial
  Intelligence (AAAI'10)}, pages 715--722. AAAI Press, 2010.

\bibitem{BCE12}
F.~Brandt, V.~Conitzer, and U.~Endriss.
\newblock Computational social choice.
\newblock In {\em Multiagent Systems}. MIT Press, 2012.

\bibitem{BCHKNSW14}
R.~Bredereck, J.~Chen, S.~Hartung, S.~Kratsch, R.~Niedermeier, O.~Such{\'y},
  and G.~J. Woeginger.
\newblock A multivariate complexity analysis of lobbying in multiple referenda.
\newblock {\em Journal of Artificial Intelligence Research}, 50:409--446, 2014.

\bibitem{BCFNN14}
R.~Bredereck, J.~Chen, A.~Nichterlein, P.~Faliszewski, and R.~Niedermeier.
\newblock Prices matter for the parameterized complexity of shift bribery.
\newblock In {\em Proceedings of the 28th Conference on Artificial Intelligence
  (AAAI'14)}, 2014.

\bibitem{BCW13}
R.~Bredereck, J.~Chen, and G.~J. Woeginger.
\newblock Are there any nicely structured preference profiles nearby?
\newblock In {\em Proceedings of the 23rd International Joint Conference on
  Artificial Intelligence (IJCAI'13)}, pages 62--68. AAAI Press, 2013.

\bibitem{BreCheWoe2013}
R.~Bredereck, J.~Chen, and G.~J. Woeginger.
\newblock A characterization of the single-crossing domain.
\newblock {\em Social Choice and Welfare}, 41(4):989--998, 2013.

\bibitem{BuCh}
L.~Bulteau and J.~Chen.
\newblock $d$-dimensional {E}uclidean preferences.
\newblock In preparation.

\bibitem{Cai03}
L.~Cai.
\newblock Parameterized complexity of vertex colouring.
\newblock {\em Discrete Applied Mathematics}, 127(3):415--429, 2003.

\bibitem{CCDF97}
L.~Cai, J.~Chen, R.~G. Downey, and M.~R. Fellows.
\newblock Advice classes of parameterized tractability.
\newblock {\em Annals of Pure and Applied Logic}, 84(1):119--138, 1997.

\bibitem{ChFaNiTa2014}
J.~Chen, P.~Faliszewski, R.~Niedermeier, and N.~Talmon.
\newblock Combinatorial voter control in elections.
\newblock In {\em Proceedings of the 39th International Symposium on
  Mathematical Foundations of Computer Science (MFCS'14)}, 2014.

\bibitem{CM08}
J.~Chen and J.~Meng.
\newblock On parameterized intractability: {H}ardness and completeness.
\newblock {\em The Computer Journal}, 51(1):39--59, 2008.

\bibitem{ChPrWo}
J.~Chen, K.~Pruhs, and G.~J. Woeginger.
\newblock Characterizations of the one-dimensional {E}uclidean domain.
\newblock In preparation.

\bibitem{CELM07}
Y.~Chevaleyre, U.~Endriss, J.~Lang, and N.~Maudet.
\newblock A short introduction to computational social choice.
\newblock In {\em Proceedings of the 39th International Conference on Current
  Trends in Theory and Practice of Computer Science (SOFSEM'07)}, volume 4362
  of {\em LNCS}, pages 51--69. Springer, 2007.

\bibitem{Con10}
V.~Conitzer.
\newblock Making decisions based on the preferences of multiple agents.
\newblock {\em Communications of the ACM}, 53(3):84--94, 2010.

\bibitem{cor-gal-spa:c:sp-width}
D.~Cornaz, L.~Galand, and O.~Spanjaard.
\newblock Bounded single-peaked width and proportional representation.
\newblock In {\em Proceedings of the 20th European Conference on Artificial
  Intelligence (ECAI'12)}, pages 270--275, 2012.

\bibitem{CGS13}
D.~Cornaz, L.~Galand, and O.~Spanjaard.
\newblock {K}emeny elections with bounded single-peaked or single-crossing
  width.
\newblock In {\em Proceedings of the 23rd International Joint Conference on
  Artificial Intelligence (IJCAI'13)}, pages 76--82. AAAI Press, 2013.

\bibitem{dC85}
M.~J. A. N.~C. de~Condorcet.
\newblock {\em Essai sur l'application de l'analyse {\`a} la probabilit{\'e}
  des d{\'e}cisions rendues {\`a} la pluralit{\'e} des voix}.
\newblock Paris: L'Imprimerie Royale, 1785.

\bibitem{Die12}
R.~Diestel.
\newblock {\em Graph Theory, 4th Edition}, volume 173 of {\em Graduate Texts in
  Mathematics}.
\newblock Springer, 2012.

\bibitem{Dod76}
C.~Dodgson.
\newblock A method of taking votes on more than two issues.
\newblock Pamphlet printed by the Clarendon Press, Oxford, and headed ``not yet
  published'', 1876.

\bibitem{DS12}
B.~Dorn and I.~Schlotter.
\newblock Multivariate complexity analysis of swap bribery.
\newblock {\em Algorithmica}, 64(1):126--151, 2012.

\bibitem{DF99}
R.~G. Downey and M.~R. Fellows.
\newblock {\em Parameterized Complexity}.
\newblock Springer Verlag, 1999.

\bibitem{DF13}
R.~G. Downey and M.~R. Fellows.
\newblock {\em Fundamentals of Parameterized Complexity}.
\newblock Springer-Verlag, 2013.

\bibitem{DT11}
R.~G. Downey and D.~M. Thilikos.
\newblock Confronting intractability via parameters.
\newblock {\em Computer Science Review}, 5(4):279--317, 2011.

\bibitem{DKNS01a}
C.~Dwork, R.~Kumar, M.~Naor, and D.~Sivakumar.
\newblock Rank aggregation methods for the web.
\newblock In {\em Proceedings of the 10th International World Wide Web
  Conference (WWW'01)}, pages 613--622. ACM, 2001.

\bibitem{DKNS01b}
C.~Dwork, R.~Kumar, M.~Naor, and D.~Sivakumar.
\newblock Rank aggregation revisited, 2001.
\newblock Manuscript.

\bibitem{EF10}
E.~Elkind and P.~Faliszewski.
\newblock Approximation algorithms for campaign management.
\newblock In {\em Proceedings of the 6th International Workshop on Internet and
  Network Economics (WINE'10)}, volume 6484 of {\em LNCS}, pages 473--482.
  Springer, 2010.

\bibitem{EFS14}
E.~Elkind, P.~Faliszewski, and P.~Skowron.
\newblock A characterization of the single-peaked single-crossing domain.
\newblock In {\em Proceedings of the 28th Conference on Artificial Intelligence
  (AAAI'14)}, 2014.

\bibitem{EFS09}
E.~Elkind, P.~Faliszewski, and A.~Slinko.
\newblock Swap bribery.
\newblock In {\em Proceedings of the 2nd International Symposium on Algorithmic
  Game Theory (SAGT'09)}, volume 5814 of {\em LNCS}, pages 299--310. Springer,
  2009.

\bibitem{EFS12}
E.~Elkind, P.~Faliszewski, and A.~M. Slinko.
\newblock Clone structures in voters' preferences.
\newblock In {\em Proceedings of the 13th ACM Conference on Electronic Commerce
  (EC'12)}, pages 496--513. ACM, 2012.

\bibitem{ELP13}
G.~Erd{\'e}lyi, M.~Lackner, and A.~Pfandler.
\newblock Computational aspects of nearly single-peaked electorates.
\newblock In {\em Proceedings of the 27th AAAI Conference on Artificial
  Intelligence (AAAI'13)}, pages 283--289. AAAI Press, 2013.

\bibitem{FHH09}
P.~Faliszewski, E.~Hemaspaandra, and L.~A. Hemaspaandra.
\newblock How hard is bribery in elections?
\newblock {\em Journal of Artificial Intelligence Research}, 35:485--532, 2009.

\bibitem{fal-hem-hem:j:cacm-survey}
P.~Faliszewski, E.~Hemaspaandra, and L.~A. Hemaspaandra.
\newblock Using complexity to protect elections.
\newblock {\em Communications of the ACM}, 53(11):74--82, 2010.

\bibitem{FHH14}
P.~Faliszewski, E.~Hemaspaandra, and L.~A. Hemaspaandra.
\newblock The complexity of manipulative attacks in nearly single-peaked
  electorates.
\newblock {\em Artificial Intelligence}, 207:69--99, 2014.

\bibitem{FHHR09b}
P.~Faliszewski, E.~Hemaspaandra, L.~A. Hemaspaandra, and J.~Rothe.
\newblock {L}lull and {C}opeland voting computationally resist bribery and
  constructive control.
\newblock {\em Journal of Artificial Intelligence Research}, 35:275--341, 2009.

\bibitem{FHHR11}
P.~Faliszewski, E.~Hemaspaandra, L.~A. Hemaspaandra, and J.~Rothe.
\newblock The shield that never was: Societies with single-peaked preferences
  are more open to manipulation and control.
\newblock {\em Information and Computation}, 209(2):89--107, 2011.

\bibitem{fal-rot:b:handbook-comsoc-control-and-bribery}
P.~Faliszewski and J.~Rothe.
\newblock Control and bribery.
\newblock In F.~Brandt, V.~Conitzer, U.~Endriss, J.~Lang, and A.~Procaccia,
  editors, {\em Handbook of Computational Social Choice}. Cambridge University
  Press.
\newblock To appear.

\bibitem{FFHMR10}
M.~R. Fellows, J.~Flum, D.~Hermelin, M.~M{\"u}ller, and F.~A. Rosamond.
\newblock W-hierarchies defined by symmetric gates.
\newblock {\em Theory of Computing Systems}, 46(2):311--339, 2010.

\bibitem{FJR13}
M.~R. Fellows, B.~M.~P. Jansen, and F.~A. Rosamond.
\newblock Towards fully multivariate algorithmics: Parameter ecology and the
  deconstruction of computational complexity.
\newblock {\em European Journal of Combinatorics}, 34(3):541--566, 2013.

\bibitem{FG06}
J.~Flum and M.~Grohe.
\newblock {\em Parameterized Complexity Theory}.
\newblock Springer Verlag, 2006.

\bibitem{FT87}
A.~Frank and {\'E}.~Tardos.
\newblock An application of simultaneous diophantine approximation in
  combinatorial optimization.
\newblock {\em Combinatorica}, 7(1):49--65, 1987.

\bibitem{GJ79}
M.~R. Garey and D.~S. Johnson.
\newblock {\em Computers and Intractability---A Guide to the Theory of
  NP-Completeness}.
\newblock W. H. Freeman and Company, 1979.

\bibitem{Goo54}
L.~A. Goodman.
\newblock On methods of amalgamation.
\newblock In R.~M. Thrall, C.~H. Coombs, and R.~L. Davis, editors, {\em
  Decision Processes}, pages 39--48. John Wiley and Sons, Inc., 1954.

\bibitem{GHN04}
J.~Guo, F.~H{\"u}ffner, and R.~Niedermeier.
\newblock A structural view on parameterizing problems: Distance from
  triviality.
\newblock In {\em Proceedings of the 1st International Workshop on
  Parameterized and Exact Computation (IWPEC'04)}, volume 3162 of {\em LNCS},
  pages 162--173. Springer, 2004.

\bibitem{GN07}
J.~Guo and R.~Niedermeier.
\newblock Invitation to data reduction and problem kernelization.
\newblock {\em ACM SIGACT News}, 38(1):31--45, 2007.

\bibitem{GNW07}
J.~Guo, R.~Niedermeier, and S.~Wernicke.
\newblock Parameterized complexity of {V}ertex {C}over variants.
\newblock {\em Theory of Computing Systems}, 41(3):501--520, 2007.

\bibitem{HHR07}
E.~Hemaspaandra, L.~Hemaspaandra, and J.~Rothe.
\newblock Anyone but him: {T}he complexity of precluding an alternative.
\newblock {\em Artificial Intelligence}, 171(5--6):255--285, 2007.

\bibitem{HHR97}
E.~Hemaspaandra, L.~A. Hemaspaandra, and J.~Rothe.
\newblock Exact analysis of {D}odgson elections: {L}ewis {C}aroll's 1876 voting
  system is complete for parallel access to {NP}.
\newblock {\em Journal of the ACM}, 44(6):806--825, 1997.

\bibitem{Kan87}
R.~Kannan.
\newblock Minkowski's convex body theorem and integer programming.
\newblock {\em Mathematics of Operations Research}, 12(3):415--440, 1987.

\bibitem{Kem59}
J.~G. Kemeny.
\newblock Mathematics without numbers.
\newblock {\em Daedalus}, 88(4):571--591, 1959.

\bibitem{Knoblauch2010}
V.~Knoblauch.
\newblock Recognizing one-dimensional {E}uclidean preference profiles.
\newblock {\em Journal of Mathematical Economics}, 46(1):1--5, 2010.

\bibitem{KurLaiPro2013}
D.~Kurokawa, J.~K. Lai, and A.~D. Procaccia.
\newblock How to cut a cake before the party ends.
\newblock In {\em Proceedings of the 27th Conference on Artificial Intelligence
  (AAAI'13)}, pages 555--561. AAAI Press, 2013.

\bibitem{Len83}
H.~W. Lenstra.
\newblock Integer programming with a fixed number of variables.
\newblock {\em Mathematics of Operations Research}, 8(4):538--548, 1983.

\bibitem{Lev75}
A.~Levenglick.
\newblock Fair and reasonable election systems.
\newblock {\em Behavioral Science}, 20(1):34--46, 1975.

\bibitem{Lin11}
A.~Lin.
\newblock The complexity of manipulating $k$-approval elections.
\newblock In {\em Proceedings of the 3rd International Conference on Agents and
  Artificial Intelligence (ICAART'11)}, pages 212--218. SciTePress, 2011.

\bibitem{FLLZ09}
H.~Liu, H.~Feng, D.~Zhu, and J.~Luan.
\newblock Parameterized computational complexity of control problems in voting
  systems.
\newblock {\em Theoretical Computer Science}, 410(27--29):2746--2753, 2009.

\bibitem{LZ10}
H.~Liu and D.~Zhu.
\newblock Parameterized complexity of control problems in maximin election.
\newblock {\em Information Processing Letters}, 110(10):383--388, 2010.

\bibitem{LZ13}
H.~Liu and D.~Zhu.
\newblock Parameterized complexity of control by voter selection in {Maximin,
  Copeland, Borda, Bucklin, and Approval} election systems.
\newblock {\em Theoretical Computer Science}, 498:115--123, 2013.

\bibitem{LMS12}
D.~Lokshtanov, N.~Misra, and S.~Saurabh.
\newblock Kernelization--{P}reprocessing with a guarantee.
\newblock In {\em The Multivariate Algorithmic Revolution and Beyond}, pages
  129--161, 2012.

\bibitem{Mar06}
D.~Marx.
\newblock Parameterized complexity and approximation algorithms.
\newblock {\em The Computer Journal}, 51(1):60--78, 2008.

\bibitem{McC06}
J.~C. McCabe-Dansted.
\newblock Approximability and computational feasibility of {D}odgson's rule.
\newblock Master's thesis, University of Auckland, 2006.

\bibitem{Nie06}
R.~Niedermeier.
\newblock {\em Invitation to Fixed-Parameter Algorithms}.
\newblock Oxford University Press, February 2006.

\bibitem{Nie10}
R.~Niedermeier.
\newblock Reflections on multivariate algorithmics and problem
  parameterization.
\newblock In {\em Proceedings of the 27th International Symposium on
  Theoretical Aspects of Computer Science (STACS'10)}, volume~5 of {\em
  LIPIcs}, pages 17--32. Schloss Dagstuhl--Leibniz-Zentrum f\"ur Informatik,
  2010.

\bibitem{PY96}
C.~H. Papadimitriou and M.~Yannakakis.
\newblock On limited nondeterminism and the complexity of the {V-C} dimension.
\newblock {\em Journal of Computer and System Sciences}, 53(2):161--170, 1996.

\bibitem{Roberts1977}
K.~W. Roberts.
\newblock Voting over income tax schedules.
\newblock {\em Journal of Public Economics}, 8(3):329--340, 1977.

\bibitem{Cake-book}
J.~Robertson and W.~Webb.
\newblock {\em Cake-Cutting Algorithms---Be Fair If You Can}.
\newblock A.K. Peters, Massachusetts, 1998.

\bibitem{SFE11}
I.~Schlotter, P.~Faliszewski, and E.~Elkind.
\newblock Campaign management under approval-driven voting rules.
\newblock In {\em Proceedings of the 25th AAAI Conference on Artificial
  Intelligence (AAAI'11)}, pages 726--731. AAAI Press, 2011.
\newblock FPT approximation schemes available in an unpublished full version of
  the paper.

\bibitem{Sim13}
N.~Simjour.
\newblock {\em Parameterized Enumeration of Neighbour Strings and {K}emeny
  Aggregations}.
\newblock PhD thesis, University of Waterloo., 2013.

\bibitem{SF13}
P.~Skowron and P.~Faliszewski.
\newblock Approximating the maxcover problem with bounded frequencies in {FPT}
  time.
\newblock Technical Report arXiv:1309.4405 [cs.DS], arXiv.org, Sept. 2013.

\bibitem{Walsh2007}
T.~Walsh.
\newblock Uncertainty in preference elicitation and aggregation.
\newblock In {\em Proceedings of the 22nd AAAI Conference on Artificial
  Intelligence (AAAI'07)}, pages 3--8. AAAI Press, 2007.

\bibitem{WYGFC13}
J.~Wang, M.~Yang, J.~Guo, Q.~Feng, and J.~Chen.
\newblock Parameterized complexity of control and bribery for $d$-approval
  elections.
\newblock In {\em Combinatorial Optimization and Applications - 7th
  International Conference (COCOA'13)}, volume 8287 of {\em LNCS}, pages
  260--271. Springer, 2013.

\bibitem{YG13}
Y.~Yang and J.~Guo.
\newblock Complexity of control behaviors in $k$-peaked elections for a variant
  of approval voting.
\newblock Manuscript, Max-Planck Institute, June 2013.
\newblock {\tt arXiv:1304.4471v3 [cs.GT]}.

\end{thebibliography}

\end{document}